\documentclass[12pt]{article}

\usepackage{fullpage,enumerate}

\usepackage{amsmath,amssymb,setspace,algpseudocode,algorithm,algorithmicx,graphicx}

\newtheorem{thm}{Theorem}
\newtheorem{lem}[thm]{Lemma}

\newtheorem{claim}{Claim}
\def\qed{\hspace*{\fill}\rule{2mm}{2mm}}
\newcommand{\proof}[1]{{\bf Proof.} #1  \hfill \qed}

\newcommand{\as}[1]{|\!|#1|\!|}
\newcommand*{\vv}[1]{\overrightarrow{#1}}
\renewcommand{\vec}[1]{\boldsymbol{#1}}

\begin{document}

\title{Dynamic Programming  Optimization in Line of Sight Networks}

\author{Pavan Sangha, Prudence W. H.  Wong, Michele Zito \\ Department of Computer Science  \\ University of Liverpool \\
Liverpool, United Kingdom \\
e-mail: {\small\texttt{pavan@liverpool.ac.uk, pwong@liverpool.ac.uk, michele@liverpool.ac.uk}}}

\maketitle

\begin{abstract}
Line of Sight (LoS) networks were designed to model wireless communication
in settings which may contain obstacles restricting node visibility. For fixed positive integer $d$, and positive integer $\omega$, a 
graph $G=(V,E)$ is a ($d$-dimensional) LoS network with range
parameter $\omega$ if it can be
embedded in a cube of side size $n$ of the $d$-dimensional integer grid so that each pair of
vertices in $V$ are adjacent if and only if their embedding
coordinates differ only in one position and such difference is less than
$\omega$. 

In this paper we investigate  a dynamic programming (DP) approach which
can be used to obtain efficient algorithmic solutions for various combinatorial problems in
LoS networks.
In particular DP solves the Maximum Independent Set (MIS) problem in
LoS networks optimally for any $\omega$ on {\em narrow} LoS
networks (i.e. networks which can be embedded in a $n \times k \times k \ldots
\times k$ region, for some fixed $k$ independent of $n$). 
In the unrestricted case it has been shown that the MIS problem is NP-hard
when $ \omega > 2$ (the hardness proof goes through  for any
$\omega=O(n^{1-\delta})$, for fixed $0<\delta<1$). 
We describe how DP can be used as a building block in the design
of good approximation algorithms.  In particular we
present a 2-approximation algorithm and a fast polynomial time
approximation scheme for the MIS problem in arbitrary
$d$-dimensional LoS networks. Finally we comment on how the approach
can be adapted to solve a number of important
optimization problems in LoS networks.\end{abstract}

%%%%%\newpage
\section{Introduction}
\label{sec:intro}

A
 wireless network typically consists of devices that communicate using
 radio
 frequencies, bluetooth or other wireless
 protocols. Geometric 
graphs often provide a good model for such networks with vertices
representing the devices, and edges associated to the communication
ability between pairs of devices.
A 
number of issues reduce the potential of wireless
communication. First of all there is typically a communication range
restriction: devices should be close in distance in order to be able
to communicate. 
Also, real world wireless networks are typically prone to line of
sight restrictions, often due to the presence of a large number of
obstacles, like those found in urban settings.
A group of devices can only communicate if they are both close 
and also there is no obstacle between them.
While the presence of obstacles can be difficult to model, it is clear that a good model of wireless network should ideally incorporate both communication range restrictions and line of sight restrictions. 

Frieze 
et al.~\cite{frieze2009line} introduced the notion of (random)
2-dimensional Line of
Sight (LoS) networks and studied connectivity problems in this setting.
Since then connectivity in higher dimensions, percolation and
communication problems have been
analyzed~\cite{farczadi2013connectivity,bollobas2009line,czumaj2007communication}
in the same model. 
For positive integers $d$, $k$ and $n$, let $\mathbb{Z}^d_{n}$ be the
$d$-dimensional cube  $\{1, \ldots, n\}^d$. We will also work with
{\em narrow} ``cubes'' $\{1, \ldots, n\}  \times  \{1, \ldots, x_1\} \times \ldots \times
  \{1, \ldots, x_d\}$, where the $x_i$ are positive integers bound by
  $k$, a positive
integer constant independent of $n$, which we denote
by $\mathbb{Z}^d_{n,k}$.  
We say that distinct points $p_{1}$ and $p_{2}$ in one of these cubes {\em share a line of
  sight} if their coordinates differ in a single place. In
this paper we mainly work with (vertex) weighted graphs: these will be
described by triples $(V,E,w)$ where as usual $V$ is the set of
nodes, $E$ is the set of edges, and $w$ is a function
assigning a positive weight to each element of $V$. An {\em
  unweighted} graph is a weighted graph whose weighting function is the
constant $w(v)=1$ for all $v \in V$. A graph $G=(V,E,w)$ is said
to be a {\em (narrow) Line of Sight (LoS) network (with parameters $n$, $k$ and
  $\omega$)} if there exists an embedding $f_{G}:V \rightarrow
\mathbb{Z}^d_{n}$ (resp. with $f_G(V) \subseteq \mathbb{Z}^d_{n,k}$)  such that $\{u,v\} \in E$ if and only if $f_{G}(u)$
and $f_{G}(v)$ share a line of sight and the (Manhattan) distance
between $f_G(u)$ and $f_{G}(v)$ is less than $\omega$. 
We refer to $\omega$ as the {\em range
  parameter} of the network. 
LoS networks keep the distance constraints of other geometric models
\cite{chiu2013stochastic} but also provide a simple mechanism to model
communication in an environment containing obstacles.

In
this work we mainly focus on the Maximum Independent Set (MIS) problem
(as defined for instance in \cite{diestel2000graph}). In fact we work with
the weighted version of this problem, where one is after an
independent set of the largest possible total weight, defined as the
sum of the weights of the elements of the chosen set.
(Narrow) unweighted LoS networks could be seen as simplistic models of urban
environment (e.g. a portion of Manhattan, where junctions correspond
to nodes and range constraints define the possible connections). In
this context large independent sets could
be used to assign police officers to junctions so as to maximize the
police presence (but still guarantee that two officers cannot shoot
each other, assuming their gun's firing range is at most $\omega-1$ blocks).
In
 general finding the largest
independent sets in a graph is NP-hard~\cite{gary1979computers} and even finding
good approximate solutions in polynomial time is
difficult~\cite{haastad1999clique}. On LoS networks, in the
unweighted case, if
$\omega=2$ or $n$ the problem can be
solved optimally in polynomial time. However, Sangha and Zito~\cite{DBLP:conf/caldam/SanghaZ17}
showed that  the general problem is NP-hard for $\omega=O(n^{1-\delta})$ where
$0<\delta<1$ is fixed, and that it admits a $d$-approximation
for any~$\omega$ and an efficient polynomial time approximation scheme
(EPTAS) ~\cite{cesati1997efficiency} for constant $\omega$. 

In
this paper we describe two algorithms that
are guaranteed to output good quality solutions for the MIS in LoS
networks when $\omega$ is a constant independent of the cube size
$n$. The first one is an approximation algorithm that 
returns a solution whose total weight is at least half the weight of an optimal
solution on any given instance, in any dimension $d$. For $d>2$ no
such algorithm was known. The second one
is a new polynomial time approximation scheme (EPTAS
\cite{cesati97ipl})  that is
faster than the one in \cite{DBLP:conf/caldam/SanghaZ17}.
%\color{red}Details in Section
%\ref{sec:ptas}. [OR SHOULD THEY BE HERE??]\normalcolor\
The
 two results hinge on a dynamic programming strategy that can be
used to solve optimally the MIS problem on narrow instances, for any $\omega$.
%
% \rem{2017.12.21}{
%
% SELLING POINTS
% \begin{itemize}
% \item General (EPTAS applicable to many problems
% \item Compare with UDGs (simpler, more effective)
% \end{itemize}
%
% SPECIAL FEATURES
% \begin{itemize}
% \item NON-robust algorithms,
% \item needs small $\omega$.
% \end{itemize}
% }
%
%
The 
technique also finds application~\cite{bellman2014mathematical} in the following scheduling problems.
Suppose that a  company manages advertisements from some $k$
clients over a long period of $n$ discrete time points.
At any time advertisements of some subset of clients are available to be aired
but the company can only select a certain number $l$ of them to advertise due to resource limitation.
In addition some ``advertisement diversity'' policy requires that
advertisements from the same client cannot be aired more than once in
a given period of $\omega-1$ time instants.
The goal of the company is to schedule the airing of these advertisements satisfying the constraints and maximising the number of advertisements aired.
This problem (which from now on will be referred to as {\sc AdsSched})
has one slight difference from the MIS problem on narrow 2-dimensional
LoS networks, in the sense that the ``proximity'' restriction only applies to one dimension (the time dimension) but not the other (the client dimension).
Nevertheless, as to be showed later, the solution we develop can be adapted to solve this problem.
Finally, we remark that the approximation strategies described in the
context of the MIS apply to a number of other optimization problems in
LoS networks. These include
Vertex Cover, Min Dominating Set, Min Edge Dominating
Set,  Max Triangle Packing, Max $H$-matching, Max Tile Salvage.

The
rest of the paper is organized as follows. After a section containing
some useful definitions, in Section~\ref{algo}, we
describe our main technical tool: a dynamic programming approach that
solves optimally the MIS problem in narrow (LoS network) instances. We
present the algorithm, a proof of correctness and a simple application
to the {\sc AdsSched} problem defined above. The remaining sections present further applications of this idea. Section
\ref{sec:eptas} describes how the dynamic programming algorithm can be
incorporated in a semi-online \cite{albers2003online} algorithm which always returns a
good quality feasible solution to the MIS problem on
narrow instances. Section \ref{sec:approx} presents the
approximation algorithm for the MIS problem in general
$d$-dimensional LoS networks, whereas Section \ref{sec:ptas} focuses
on the EPTAS for the 2-dimensional case, and some
additional applications. Section 
\ref{sec:conclusions} wraps up the paper with a summary of the results
presented and some directions for future work.

%%%%%\newpage 
\section{Problem Definitions and Preliminaries}\label{prelim}

In this paper arrays will be $d$-dimensional tables of non-negative
numbers. In
particular, for fixed $k>0$, {\em narrow} arrays are tables of size $x_1 \times \ldots
\times x_{d-1} \times y$ where the $x_i \in \{1, \ldots, k\}$ for all
$i \in \{1, \ldots, d-1\}$ whereas $y$ is a positive integer 
no larger than $n$. It will be convenient to group the first $d-1$
indices and so we will often write $\vec{x}^{d-1} \times y$ instead of  $x_1 \times \ldots
\times x_{d-1} \times y$ or $A[\vec{i},j]$ instead of $A[i_1, \ldots,
i_{d-1}, j]$. In this context the $j$th {\em column} of array $A$ will be the collection of
elements $A[\vec{i},j]$ for all possible values of $\vec{i}$.

\begin{itemize}
\item For any array $A$ and $j_1, j_2 \in  \{1, \ldots, n\}$ with $j_1
  \leq j_2$, denote by $A[j_1 : j_2]$  the sub-array
  containing columns $j_1, \ldots, j_2$. When $j_1 = j_2  = j$ we use
  $A[j]$ instead of $A[j : j]$ unless ambiguity arises.
\item For any two arrays $A_1, A_2$ of size $\vec{x}^{d-1} \times y$, 
  we say that $A_1$ {\em agrees with} $A_2$, denoted by $A_1 \leq_a A_2$,
  if $A_1[\vec{i},j] \leq A_2[\vec{i},j]$ for all $\vec{i}$ and $j$.
\item For any array $A$, we denote by $h(A)$ (resp. $t(A)$) the {\em
    head} (resp. {\em tail}) {\em subarray} of $A$ containing all but
  the last column (resp. all but the first column) of $A$. In other words, $h(A)$ and $t(A)$ have $y-1$ columns if $A$ has $y$ columns.
\item we say that  $A_1$ is \emph{consistent} with $A_2$ (in symbols $A_1
  \vDash A_2$) if $t(A_1)$
  is the same as $h(A_2)$.
\item Let the {\em column sum of an array $A$ at column $j$} be the quantity
  \[\as{A}_j = \sum_{\vec{i}} A[\vec{i},j].\]
We refer to the quantity $\sum_j \as{A}_j$ as the {\em array
  sum of $A$}, and we denote it by $\as{A}$. 
\end{itemize}

\begin{figure}[tb]
\centering 
\includegraphics[scale=.35]{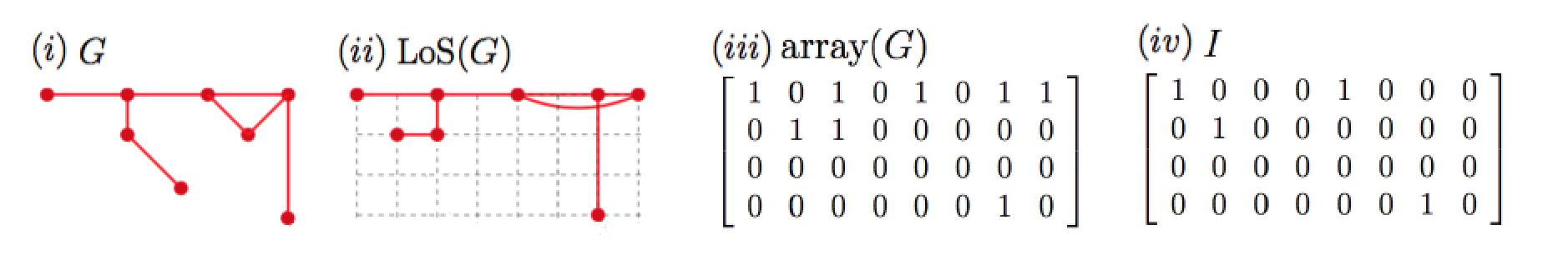}
\caption{\label{fig1} Figure (i) is a graph $G$ and Figure (ii) is its
  LoS embedding in an $8 \times 4$ rectangle in $\mathbb{Z}^2$ with $\omega=4$. Figure (iii)
  represents the array layout of $G$ (ignoring the first $\omega$ columns of
  zeroes) and Figure(iv) is an independent array of largest array sum, corresponding to the largest independent set in the graph $G$.}
\end{figure}

Given a weighted narrow LoS network $G = (V,E,w)$, let array$(G)$ be a $\vec{k}^{d-1} \times
(n + \omega)$ array satisfying array$(G)[\vec{i},j]=w(v)$ (resp ``0'') if and only if
location $(j,\vec{i}) \in \mathbb{Z}_{n,\vec{k}}$ corresponds (resp. does
not correspond) to a
vertex $v \in V$ in the LoS embedding of $G$. Also, array$(G)[\vec{i},j]  = 0$ for
any $\vec{i}$ and   $j
\in \{-(\omega-1), \ldots, 0\}$. Figure~\ref{fig1} provides an
example. In this setting an {\em independent array} $I$ of $\mathrm{array}(G)$ is any
array of size $\vec{k}^{d-1} \times (n + \omega)$ satisfying 
\begin{enumerate} 
\item $I \leq_a \mathrm{array}(G)$ and 
\item for distinct columns $j_{1},j_{2}$ if $I[\vec{i},j_{1}]>0$ and
$I[\vec{i},j_{2}]>0$ then $|j_{1}-j_{2}|\geq \omega$ and for
distinct rows indexed by $\vec{i}_{1}$ and $\vec{i}_{2}$ if $I[\vec{i}_{1},j]>0$
and $I[\vec{i}_{2},j]>0$ then either $\vec{i}_{1}$ and
$\vec{i}_{2}$ do not share a line of sight  or they do but the gap
between the values of the differing co-ordinate is at least $\omega$.
\end{enumerate}

A {\em feasible array} 
$W$ is an array of size $\vec{k}^{d-1} \times
\omega$ only containing zeros or ones and such that there exists a $\vec{k}^{d-1} \times
\omega$  independent array $I_W$ such that $W[\vec{i},j]=1$ (resp $=0$)
if and only if $I_W[\vec{i},j]>0$ ($=0$). The array $I_W$ is a {\em
  witness of} $W$. Since any feasible array has
exactly $\omega$ columns it contains at most one non-zero entry per
column.
%, but also at most one non-zero entry  per row. 
We denote by
$\mathcal{F}$ the set of all feasible arrays of size $\vec{k}^{d-1}  \times
\omega$, and for each $j \in \{1, \ldots, n\}$,  
$\mathcal{F}_{G,j} \subseteq \mathcal{F}$ is the set of feasible
arrays $W$ satisfying $W \leq_{a}$ array$(G)[j-\omega+1 : j]$. Note that in
particular for any independent array $I$ of array$(G)[-(\omega-1) : j]$ for
$1\leq j \leq n$ is the witness of some $W \in \mathcal{F}_{G,j}$.

We observe that $\mathcal{I}$ is an independent set of $G$ if and only
if ${\cal I}$ is an independent array of array$(G)$. 
Thus finding a maximum total weight independent set in $G$ is
equivalent to finding the independent array of array$(G)$ with the
largest array sum (we refer to such an array as a {\em largest
  independent array}). In Section~\ref{algo} we show how a simple DP
algorithm finds an independent array of $\mathrm{array}(G)$
with the largest array sum by working with the feasible arrays of $\mathrm{array}(G)$.
Because of this correspondence, in the next section 
we refer to $\mathrm{array}(G)$ as $G$ and we work with arrays instead
of graphs.

%%%%%\newpage 
\section{Dynamic Programming} \label{algo}

\begin{algorithm}[htb] 
{\footnotesize 
  \begin{algorithmic}[1] 
\State /* Initialisations */
   \State MIS$[\vec{0},\vv{0}]=0$, where
   $\vv{0}$ is the $\vec{k}^{d-1} \times \omega$ array of all $0$'s. 
   \For {$j=1,\ldots,n$}
    \For {$W \in \mathcal{F}$} 
\State MIS$[j,W] = 0$
\EndFor
\EndFor
\State
\State /* Array sums computation */
   \For {$j=1,\ldots,n$} \label{10}
    \For {$W \in \mathcal{F}_{G,j}$} 
    \State let $W^*$ be the feasible array in $\mathcal{F}_{G,j-1}$,
    $W^* \vDash W$, maximizing MIS$[j-1,W^*]$
      \State MIS$[j,W]= W[\omega]^{T}\cdot G[j] +$ MIS$[j-1,W^*]$   \label{key}
\State pred$[j,W] = W^*$
    \EndFor
   \EndFor \label{16}
\State
\State /* Retrieving the independent set */
\State Find $W^* \in {\cal F}_{G,n}$ that maximizes MIS$[n,W]$
\State Set $I$ as the rightmost column of $W^*$
\For {$j=n$ {\bf downto} 2}
\State $W^* = $ pred$[j,W^*]$
\State Redefine $I$ as the rightmost column
  of pred$[j,W^*]$ concatenated with $I$ 
\EndFor
\State {\bf return} $I$
  \caption{Computing the largest independent array in $G$}
	\label{dpalgo}
  \end{algorithmic}
} 
\end{algorithm}

\begin{figure}[t]
\includegraphics[scale=.4]{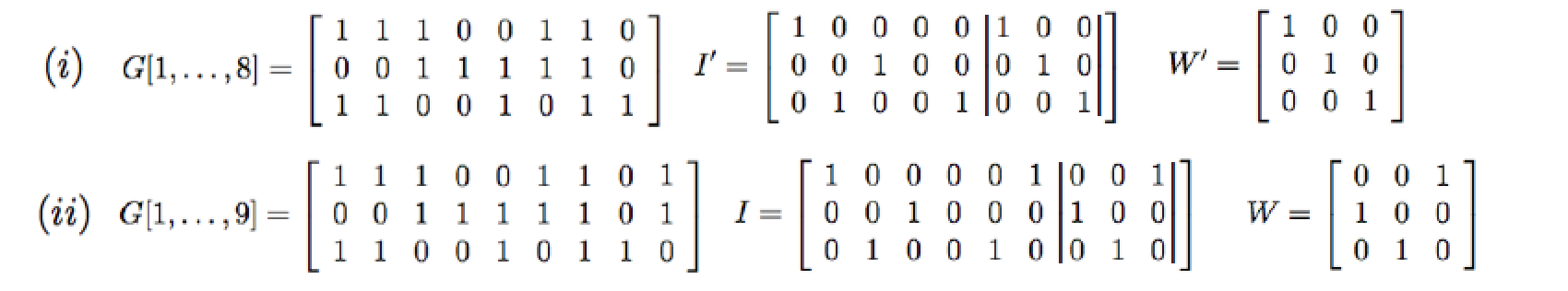}
\centering
\caption{Figure (i) shows 8 columns of an array $G$ and the
  independent array $I'$ of $G[1 : 8]$ with the largest array sum
  satisfying $I'[6 : 8] \equiv W'$.  In Figure (ii) the independent
  array $I$ is the independent array of $G[1 : 9]$ which has the
  largest array sum satisfying $I[7 : 9]\equiv W$. Note $W' \vDash W$ and that $I$ can be obtained from $I'$ by appending the last column of $W$ to $I'$.}
\label{fig2} 
\end{figure}

Given 
the array $G$ of size $\vec{k}^{d-1}  \times
(n + \omega)$, the main idea of the optimal algorithm we describe in 
this section is to be guided in its choices by a table containing 
array sums of independent arrays of portions of $G$. For each $j \in
\{0, \ldots, n\}$, the process manages a table $\mathrm{MIS}[j,W]$, indexed by $j$ as well as all
possible $\vec{k}^{d-1}  \times \omega$ feasible arrays $W$. 
If $W \in \mathcal{F}_{G,j}$, 
we try to extend the independent arrays in $G[-(\omega-1) : j-1]$ to
independent arrays in $G[-(\omega-1) : j]$ witnessing $W$. Let $I'$ be an independent
array in $G[-(\omega-1) : j-1]$ such that $I'[j-\omega : j-1]$ is a
witness 
for some $W' \in \mathcal{F}_{G,j-1}$ and assume that  $W'$ is consistent with
$W$. 
By considering the next column of $G$, we extend $I'$ to an
independent array $I$ of $G[-(\omega-1) : j]$ which is a witness to
$W$. MIS$[j,W]$ contains the array sum of an independent set whose
right-most $\omega$ columns are witnessed by $W$. The expression 
$W[\omega]^{T}\cdot G[j]$ on line \ref{key} of Algorithm \ref{dpalgo} is the Frobenius product
of $W[\omega]$ and $G[j]$.
Figure~\ref{fig2} shows an example in the two
dimensional case.  
Array pred$[j,W]$ keeps track of the extension that maximizes the size
of $I'$.

Once this is completed for all $j$'s the information in the array pred
can be used to retrieve an actual independent set. The following
result summarizes the computational properties of Algorithm \ref{dpalgo}.

\begin{thm}
\label{DPmain}
Algorithm~\ref{dpalgo} computes a maximum independent set of a
weighted narrow
LoS network $G$ in time $O(n (k^{(d-1)/\omega}\ \omega)^{k^{d-1}})$.
%(resp. $O(n\ k^{d-1}\ \omega^{k^{d-1}})$).
\end{thm}

\proof{
The 
 proof is by a simple reductio ad absurdum for
  each $j$  (similar, say, to the one described in \cite[Theorem
  15.1]{cormen09:_introd_algor} in the context of the longest common
  subsequence problem). 
Denote  by $I$ an independent array of $G$ of maximum array sum. 
There must be an independent array $I'$ of $G[-(\omega-1) : n-1]$
such that:
\[\as{I} = \as{I}_{\omega+n}+ \as{I'}.\]
Furthermore $t(I'[n-\omega : n-1]) = h(I[n-\omega+1 : n])$, in other
words the two independent arrays must be consistent. But then we are safe
to assume that 
\[\as{I'} = \max_{W' \in \mathcal{F}_{G,n-1}: W'
        \vDash W} \mathrm{MIS}[n-1,W'] )\]
(where $W$ is witnessed by $I[n-\omega+1 : n]$) for otherwise replacing $I'$ by the independent set on the right
hand side would give us a larger set for $I$ contradicting its
optimality. By the same token, $I[n-\omega+1 : n]$ must be a witness
of the feasible array $W$
maximizing $W[\omega]^{T}\cdot G[j] + \max_{W' \in \mathcal{F}_{G,n-1}: W'
        \vDash W } \mathrm{MIS}(n-1,W')$. Therefore the independent
      set returned by Algorithm 1 is at least as large as $I$.

For each 
$j$, $|\mathcal{F}_{G,j}| = O(\omega^{k^{d-1}})$ as the elements of this set
are $\vec{k}^{d-1}  \times \omega$ tables with at most a single non-zero entry in
each row\footnote{Each row can be filled in at most $\omega+1$ ways,
  and there is $k^{d-1}$ rows: at most  $(w+1)^{k^{d-1}}$ possibilities. Assuming
  also each column has at most one non-zero entry we get $\binom{k^{d-1}}{j}\binom{\omega}{j} \times j!$ if there is $j \in \{0, \ldots, k^{d-1}\}$
  non-empty rows (and columns). }. 
Let  $t=\left\lceil\frac{k}{\omega} \right\rceil$. Each column of a
feasible array $W$ is a $d-1$ dimensional cube of side length $k$. The maximum number
of non-zero elements in dimension one, say, is $t$. Furthermore the
cube is a $d-1$ dimensional object, hence an obvious upper bound on
the number of non-zero elements of $I$ is 
\[t \times k^{d-2}.\]
Finally, the maximum number
  of elements of ${\cal F}_{G,j}$ that are consistent with a given $W
  \in {\cal F}_{G,j-1}$, for each $j$, is at most
\[\binom{k^{d-1}}{t \cdot k^{d-2}}.\] 
To see this notice that, starting
    from an arbitrary $W$, we get an element of $ {\cal F}_{G,j-1}$ by chopping off
    the first column of $W$ and adding an extra column at the other
    end. 
A column is a $d-1$ dimensional cube with $k^{d-1}$ positions and
there's at most $t \cdot k^{d-2}$
non-zero positions to be placed in that.

Back to Algorithm \ref{dpalgo}, by  the counting argument above, the loop between line \ref{10} and \ref{16}
can be completed in time $O(n\ (k^{(d-1)/\omega}\ \omega)^{k^{d-1}})$
and the result follows. 
} %proof

\paragraph{Extensions} The DP algorithm described in this section can
be adapted to solve optimally a host of other optimization problems in
narrow LoS networks. The smallest vertex covers or dominating sets,
the largest triangle
packings, or $H$-matchings  and many other ``hard'' combinatorial 
structures can all be
found  in polynomial time by using obvious modifications of the strategy described above. Here we show that even problems of a
slightly different nature can be solved optimally by our DP
approach. An instance of the {\sc AdsSched} problem defined in Section
\ref{sec:intro} can be encoded
by an array $G$ exactly like the MIS in LoS networks. The only
difference is in the definition of feasible solution. Therefore Algorithm \ref{dpalgo} can also be used to solve the
{\sc AdsSched} problems, provided  the definition of ${\cal
  F}_{G,j}$ is slightly modified. In this case the elements of this set are $k \times \omega$ arrays 
$W$ satisfying the following conditions
\begin{enumerate}[(i)]
\item $W \leq_a G[j-\omega+1 : j]$
\item $W$ contains at most one non-zero element in each row.
\item $W$ contains at most $l$ non-zero elements in each column.
\end{enumerate}

\begin{thm}
\label{schedextension}
Algorithm~\ref{dpalgo} solves {\sc AdsSched} optimally in time $O(n\ k^{l} \ \omega^{k})$.
\end{thm}
\proof{
The correctness of the process follows from that of Algorithm
\ref{dpalgo} as proved in Theorem \ref{DPmain}.
  As to the running time, the only difference is in the maximum number
  of elements of ${\cal F}_{G,j}$ that are consistent with a given $W
  \in {\cal F}_{G,j-1}$, for each $j$: there is at most $\binom{k}{l}$
  of them. The result follows. 
} %proof

%%%%%\newpage 
\section{Semi-online Approximation Algorithms}
\label{sec:eptas}

The DP algorithm in Section~\ref{algo}
solves optimally the offline version of the MIS problem in
narrow LoS networks and several other related problems, where the entire input is known in advance. 
This is unrealistic in various practical settings. For example if 
the time parameter $n$ in the scheduling problem is large, possibly spanning a year or more,
then it is likely the input evolves over time. 
In such sitations it may be desirable to take a different approach,
aiming for {\em online} algorithmic solutions with good performance
guarantees. 
In this section we consider semi-online strategies that are allowed to
observe the input up to a certain look-ahead distance. We show that we
can achieve $(1+\epsilon)$-approximation with a look-ahead distance
dependent on $\epsilon>0$.
The quality of the approximation can be traded-off against the algorithm
running time as well as how much look-ahead it is allowed. We state
our main result in terms of the MIS problem, but the strategy can be
applied to any of the optimization problems described at the end of
Section \ref{algo}.

\begin{thm}
\label{semi}
 There is a semi-online algorithm that for any $\epsilon>0$ computes a feasible solution
 for the {\rm MIS} problem in a narrow LoS network in dimension $d$ that is a
 $(1+\epsilon)$-approximation of the optimum, in time
%\color{red}
\[O\left(\left(1 + \frac{1}{\epsilon}\right) n\  k^{d-1} \
  (k^{(d-1)/\omega}\ \omega)^{k^{d-1}}\right).\]
%\normalcolor
\end{thm}

The main idea of the algorithm mentioned in Theorem \ref{semi} is similar to that of the EPTAS
described in \cite{DBLP:conf/caldam/SanghaZ17} for 
general LoS networks. We will argue however
that the process we describe here is much faster than the algorithm in
that paper, provided $k$ is a constant independent of $n$.

Let $G[j_1:j_2]$ where $j_1<j_2$ describe the subgraph of the narrow LoS
network $G$ consisting of vertices which are embedded in the region  
\[\{j_1, \ldots, j_2\}  \times  \underbrace{\{1, \ldots, k\} \times \ldots \times
  \{1, \ldots, k\}}_{d-1\ \mbox{\scriptsize times}}\]
and their induced edges. Let $I_{r}$ denote a maximum independent
set in $G[1:r\omega]$. A {\em phase} in the algorithm starts by computing
$I_{0}$ in a subgraph of $G$ consisting of some column $j_0$ and
proceeds to compute $I_{r}$, for $r \geq 1$, in
$G[j_0 : j_0+ r\omega -1]$ provided $|I_{r}| \geq (1+\epsilon)
|I_{r-1}|$. Thus each $I_{r}$ in the sequence satisfies $|I_{r}| \geq
(1+\epsilon)^{r}|I_{0}|$. In addition, using the structural properties
of a LoS network embedding, we may infer that $|I_{r}| \leq
k^{d-1} r$: at most $r$ vertices can be added to $I_r$ in every row,
and there is $k^{d-1}$ rows altogether.

Let
$r^{*}$ be the least $r$ for which  
\begin{equation}
  \label{eq:2}
|I_{r^{*}+1}|<(1+\epsilon)|I_{r^{*}}|  
\end{equation}
We refer to this as the {\em stopping point} of the current
phase. When condition (\ref{eq:2}) is reached the process starts
another phase from $j_0 = (r^*+1) \omega$.

We start our analysis by proving an upper bound on $r^*$.

%%%\newpage
\begin{lem}
\label{eptas1} 
In each phase of the algorithm
\[r^{*} \leq \left(1+\frac{1}{\epsilon} \right) \frac{k^{d-1}}{(\log 2)^2}.\]
\end{lem}
\proof{
Throughout a phase we have $|I_{r}| \geq
(1+\epsilon)^{r}|I_{0}|$. 
Thus $r^*$ is bounded above by the smallest
$r$ for which 
\[k^{d-1} r < (1+\epsilon)^{r}.\]
Such number is not larger than the smallest $r$ satisfying
\[{\rm e}^{\sqrt{k^{d-1} r}} < (1+\epsilon)^{r}.\]
 Taking the logarithms, this is equivalent to
\begin{equation}
\sqrt{k^{d-1} r} < r \log (1+\epsilon).
\label{best-bound}
\end{equation}
Assuming $\epsilon < c$ for some $c>0$ this inequality holds if
\[\sqrt{k^{d-1} r} < r\  C\  \epsilon \quad \Leftrightarrow \quad  r >
\frac{k^{d-1}}{(C \epsilon)^2}.\]
(here $C = \frac{\log (1+c)}{c}$).
For $\epsilon \geq c$ inequality (\ref{best-bound}) is satisfied if
\[\sqrt{k^{d-1} r} < r \log (1+c)\]
which is equivalent to
\[r > \frac{k^{d-1}}{(\log (1+c))^2}.\] 
The lemma follows using $c=1$. 
} % proof

We may now complete the proof of Theorem \ref{semi}.
To obtain a $(1+\epsilon)$-approximation to the maximum independent
set once $r^{*}$ is obtained we remove $G_{r^*+1}$ from the graph $G$
and apply the procedure iteratively to the graph
$\overline{G_{r^*+1}}$. Arguing as in
\cite{DBLP:conf/caldam/SanghaZ17}  we have that if $I'$ is the
independent set obtained from applying the procedure to
$\overline{G_{r^*+1}}$ then $I_{r^{*}} \cup I'$ is a
$(1+\epsilon)$-approximation to the maximum independent set in $G$.

As to the running time, computing $I_{r^*}$ in each phase takes time
$O(r^*\ \omega\ (k^{(d-1)/\omega}\ \omega)^{k^{d-1}})$ by Theorem \ref{DPmain}. Using the bounds on $r^*$
from Lemma \ref{eptas1}  this can be rewritten as 
%\color{red}
\[O\left(\left(1+\frac{1}{\epsilon}\right) k^{d-1} \omega\
  (k^{(d-1)/\omega}\ \omega)^{k^{d-1}}\right).\] 
%\normalcolor
Finally since there are at most $n/\omega$ phases  (we always remove at least
$\omega$ columns of $G$ in each phase) the algorithm has a
worst-case running time of
%\color{red}
\[O\left(\left(1+\frac{1}{\epsilon}\right) n\ k^{d-1}
  (k^{(d-1)/\omega}\ \omega)^{k^{d-1}}\right).\]
%\normalcolor

\paragraph{Additional Remarks} 
The same framework can be used to devise semi-online $(1+\epsilon)$-approximation
heuristics for finding smallest vertex covers or dominating sets,
the largest triangle
packings, or $H$-matchings. The
analysis of the approximation performance is largely unchanged. The run-time in
each case is affected by the run time of
the specific DP algorithm used to complete each phase. 
The
approximation heuristics are semi-online since at any moment in time
we never work on more than $r^* \omega$ columns of
the input data.

%%%%%\newpage 
\section{Approximation Algorithms for Unrestricted LoS Networks}
\label{sec:approx}

In this section we show how the DP approach described in Section
\ref{algo} can be exploited to define an approximation algorithm for
the MIS in arbitrary $d$-dimensional LoS networks, for $d \geq 2$. To
avoid cluttering the presentation we
first describe and analyze the algorithm for the special case
$d=2$. Then we outline the modifications necessary to extend the
algorithm to the general $d$-dimensional case.

In 2-dimensions, we are given a LoS network (embedded in
$\mathbb{Z}^2_{n}$) with range parameter
$\omega>2$. The main idea is to split the input data into {\em
  strips}, each being a narrow LoS network,
apply the DP algorithm to each strip, and then combine the solutions
obtained for the strips into a solution for the whole instance.

In what follow let $k = \omega - 1$ and let $G_{[i]}$ be the {\em
  strip} formed by the vertices of $G$ embedded in rows $ki+1, \ldots, 
k(i+1)$ of $\mathbb{Z}^2_{n}$, for $i \in \{0, \ldots, n/k-1\}$  (assume
$k$ divides $n$ for simplicity). Parameter $i$ in this context is the
{\em strip index}.

\begin{figure}[htb]
  \centering
  \includegraphics[scale=0.6]{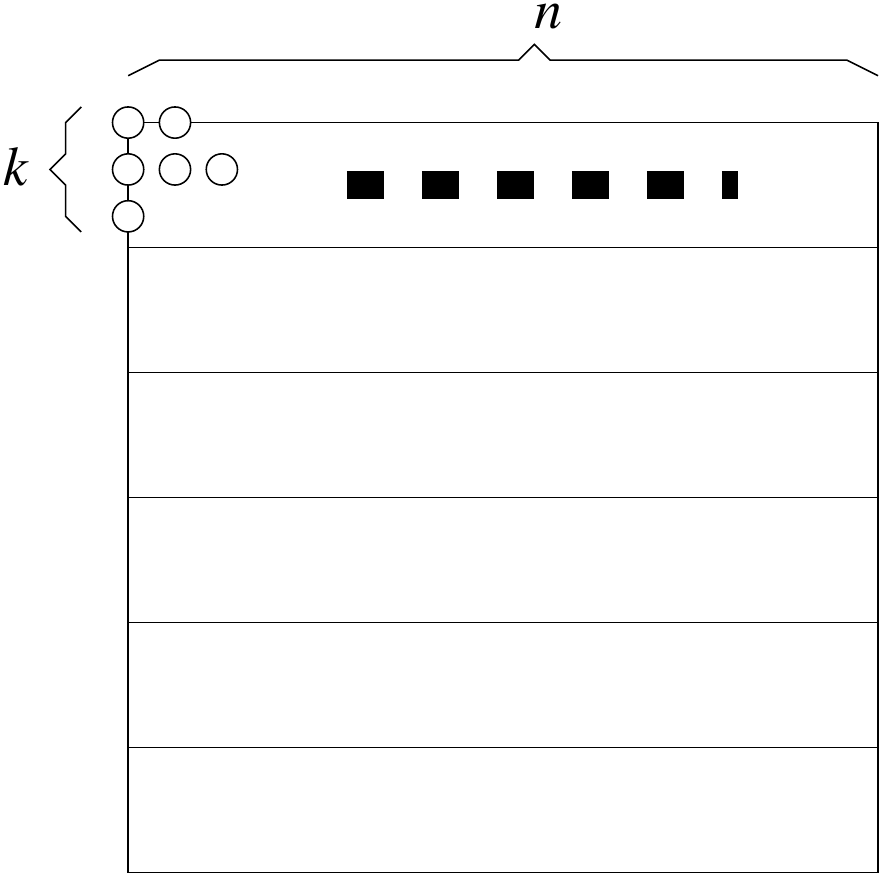}
  \caption{Splitting $G$ into $n/k$ strips ($k=3$, in the
    given example, array cells represented as small circles).}
  \label{strips}
\end{figure}

Clearly two nodes from different strips
with indices having the same parity cannot be adjacent as even if they
share a line of sight they are far from each other. Therefore the union of
a collection of 
independent sets found in all odd (resp. even) indexed  strips is an
independent set of the whole network. The sought approximation
algorithm, which we call StripIndependentSet, returns $H$, the largest
of these two sets. 

\begin{thm}
\label{d2}
For any fixed $\omega$ independent of $n$, {\rm StripIndependentSet} is a 2-approximation algorithm for the {\sc
  MIS} in a 2-dimensional LoS network.
\end{thm}
\proof{
Let's call Odd (Even) the
collection of all odd (even) indexed strips. We can use dynamic programming to find an
optimal independent set in each strip. Let DP(Odd) (resp. DP(Even)) be
the independent set found using Algorithm 1 on each Odd
(resp Even) strip. Let ${\cal I}$ be an independent set of
maximum size in the whole graph. ${\cal I} \cap$ Odd is an independent set
of Odd so it must be
\begin{center}
  $|$DP(Odd)$| \geq |{\cal I} \cap$ Odd$|$
\end{center}
and
\begin{center}
  $|$DP(Even)$| \geq |{\cal I} \cap$ Even$|$
\end{center}
Hence
\begin{center}
  $|{\cal I}| = (|{\cal I} \cap$ Odd$|) + ({\cal I} \cap$ Even$|) \leq
  |$DP(Odd)$| + |$DP(Even)$| \leq 2 \cdot |H|$. 
\end{center} 
The process requires $O(n/k)$ DP computations, each running in time 
$O(n\ k\ \omega^{k})$ (this comes from Theorem
\ref{DPmain} substituting $d=2$). The overall run time is therefore $O(n^2\
\omega^{\omega-1})$. 
} % proof

\begin{figure}[htb]
  \centering
  \includegraphics[scale=0.6]{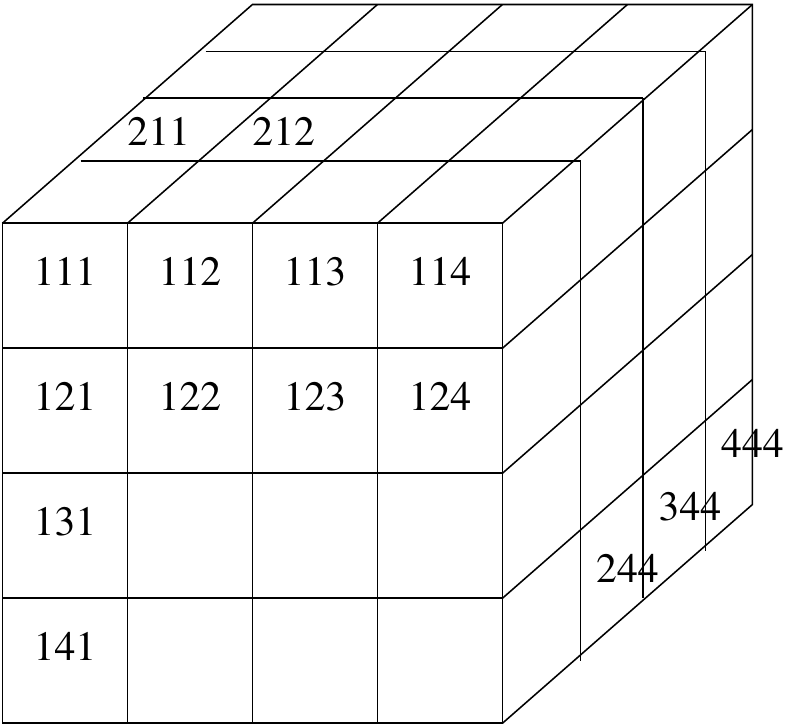}
  \caption{Splitting $G$ into strips. Here is one of the $n$
    3-dimensional ``bases'' (each small cube is labelled by the
    corresponding triple $(i_1,i_2,i_3)$). For $d=2$, one dimension is split into
    $n/k$ intervals. For $d=4$, three dimensions are split into
    $(n/k)^3$ cubes of side size $k$. DP gives an optimal solution in
    each $k^3 \times n$ strip.}
  \label{strips2}
\end{figure}

\paragraph{Generalization to $d$ dimensions}
The corner greedy strategy in
\cite{DBLP:conf/caldam/SanghaZ17} already provides a 2-approximation
algorithm for the MIS problem in 2-dimensional LoS networks. The main
advantage of the approach described above lies in the fact that
algorithm StripIndependentSet can be generalized to arbitrary
dimension $d>2$. The general strategy is unchanged but the notion of
Odd (resp. Even) strip is slightly more elaborate. As in Section
\ref{algo}, it is convenient to
think of the network
nodes as the elements of a
$d$-dimensional table.  In this context a {\em strip} is
a collection of elements 
\[G[k (i_1-1)+j_1,
\ldots, k (i_{d-1}-1)+j_d,i_d]\]
where $j_h \in \{1, \ldots, k\}$ and the vector $(i_1, \ldots, i_{d-1})$
satisfies $i_h \in \{1, \ldots,
n/k\}$, for $h \in \{1, \ldots, d-1\}$ (whereas $i_d \in \{1, \ldots, n\}$).
The vector $\vec{i} = (i_1, \ldots, i_{d-1})$ is the strip {\em index}.
A vertex belongs to an {\em odd} (resp. {\em even}) strip if its
strip index satisfies:
\[\sum_{h=1}^{d-1} i_h \mod 2 = 1 \qquad (\mbox{resp. } 0).\]
Figure
\ref{strips2} attempts to give an idea of the partitioning for $d=4$. 

\begin{claim}
Vertices belonging to different strips whose indices have the same
parity are not connected by an edge in $G$.
\end{claim}

It follows from the claim above that algorithm StripIndependentSet
returns an independent set in any LoS network embedded in $d$
dimensions and the following result complete our argument.

\begin{thm}
\label{dd}
For any fixed $\omega$ and $d$ independent of $n$, {\rm StripIndependentSet} is a 2-approximation algorithm for the {\sc
  MIS} in a $d$-dimensional LoS network.  
\end{thm}
\proof{
  The same argument used to prove Theorem \ref{d2} applies. This time
$O(n/k^{d-1})$ DP computations are needed and  each of them requires time
$O(n\ k^{(d-1)k^{d-2}}\ \omega^{k^{d-1}})$. Therefore the running time
of the process is:
\[O(n^2\ k^{(d-1)k^{d-2}-(d-1)}\ \omega^{k^{d-1}})\]
 Since $k=\omega-1$, the
overall algorithm running time is 
\[O(n^2\  (\omega-1)^{(d-1)((\omega-1)^{d-2}-1)}\
\omega^{(\omega-1)^{d-1}}).\] 
} % proof

%%%%%\newpage 
\section{Polynomial Time Approximation Schemes}
\label{sec:ptas}

The DP approach in Section \ref{algo} can also be exploited to obtain
an EPTAS for the MIS problem in general LoS networks, for any $d \geq
2$.  As in the previous section we first present the idea for the case
$d=2$.

The algorithm works on the given network
 (which is provided  with its embedding in $\mathbb{Z}^2_{n}$)
 decomposed into strips, as in Figure \ref{strips}, however we need one
additional concept. Let $h$ be a positive integer. Its value will be
fixed later on in our analysis, but for now we require that $h$ be
a fixed constant independent of $n$. A {\em block} is a collection of contiguous
strips (note that the number of rows in each block is a multiple of
$k$). For each $i  \in \{0, \ldots, h\}$, let ${\cal B}_{h,i}$ be the
partition of $G$ 
into blocks such that
the top one contains $i \times k$ rows, and all the others (except
perhaps the last one) contain $h \times k$ rows. Successive blocks are
separated by a single strip. Let $\partial {\cal B}_{h,i}$ be the
union of these ``excluded'' strips. Let $B$ be an arbitrary 
block of ${\cal B}_{h,i}$. Since the product $h \times k$ is
independent of $n$, a maximum independent set ${\cal I}_B$ in
$B$  can be found in polynomial time using Algorithm 1. The set
\[{\cal I}_i = \bigcup_{B \in {\cal B}_{h,i}} {\cal I}_B\]
is a maximum independent set of ${\cal B}_{h,i}$.
The algorithm returns the largest among ${\cal
  I}_0, {\cal
  I}_1, \ldots, {\cal
  I}_h$. Let's call ${\cal U}$ such set.
\begin{figure}[htb]
  \centering
  \includegraphics[scale=0.4]{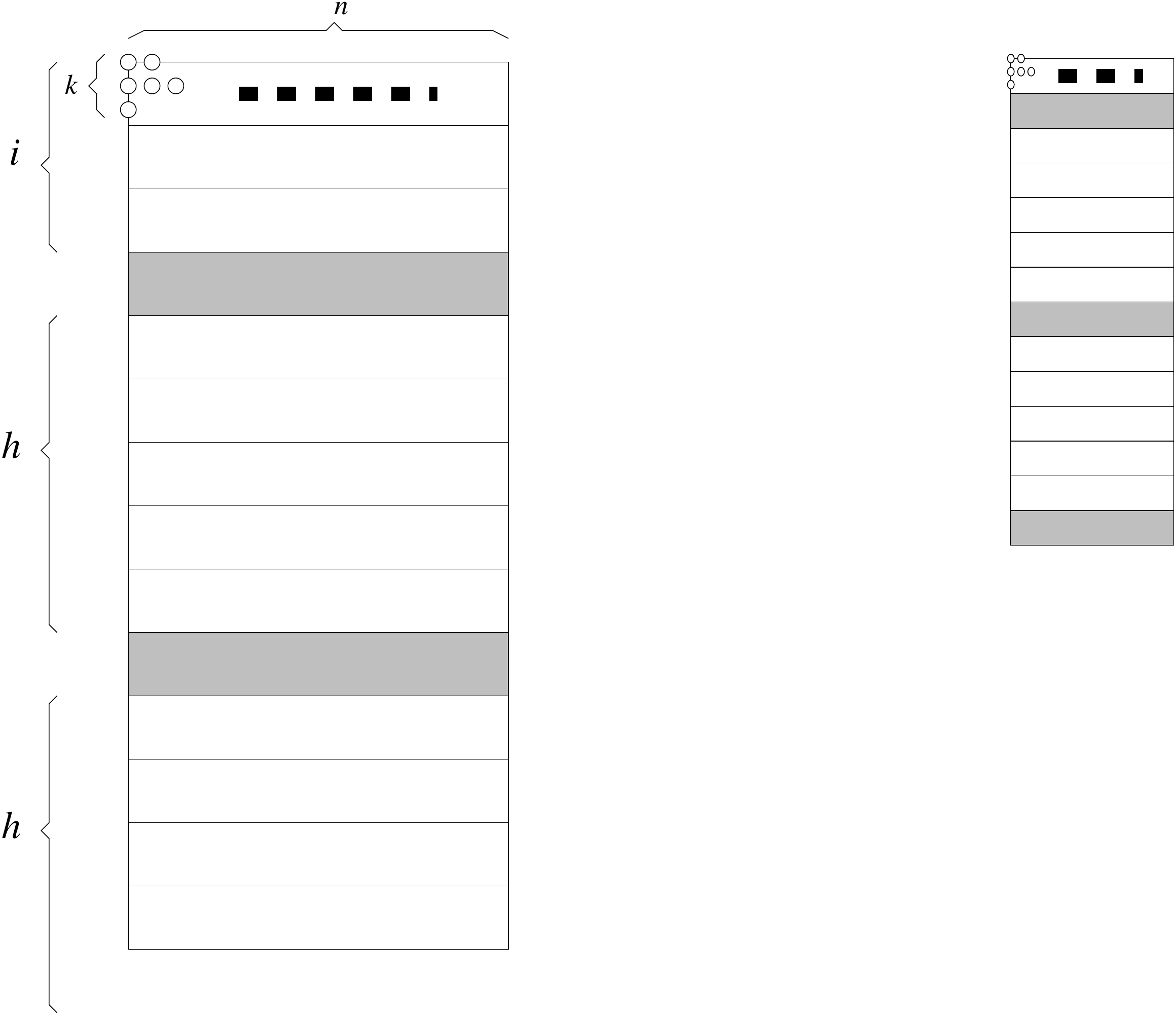}
  \caption{Splitting $G$ into blocks. The larger picture on the left
    hand side describes the
    top part of ${\cal B}_{5,3}$. The smaller picture on the right presents a
    similar schematic representation of ${\cal B}_{5,1}$. In both
    cases the greyed strips belong to the union of the excluded strips
  (i.e.  $\partial {\cal B}_{5,3}$  and $\partial {\cal B}_{5,1}$ , respectively).}
  \label{strips-ptas}
\end{figure}
Let ${\cal I}$ be a maximum independent set of the whole network.
A key property of independent sets is that a maximum independent set
in any strip $S$ of $G$ cannot have less than  $|{\cal I} \cap
V(S)|$ vertices (as the vertices in each strip in isolation are less
constrained than when they are considered as part of the whole
graph). Also, we can  write
\[|{\cal I}| = \sum_{i = 0}^h |{\cal I} \cap V(\partial {\cal
  B}_{h,i})|.\] 
But then by a simple counting argument there must be $\hat{\i} \in \{0, \ldots,
h\}$ such that $|{\cal I} \cap V(\partial{\cal B}_{h,\hat{\i}})| \leq
|{\cal I}|/(1+h)$. 
This implies that a maximum independent set ${\cal
  I}_{\hat{\i}}$ in 
${\cal B}_{h,\hat{\i}}$ (which will be eventually found by the algorithm) must satisfy
\[|{\cal I}_{\hat{\i}}| \geq |{\cal I} \cap V({\cal B}_{h,\hat{\i}})|
= |{\cal I}| - |{\cal I} \cap V(\partial {\cal B}_{h,\hat{\i}})| >
\frac{h}{1+h}|{\cal I}|.\]
Thus we have
\[\frac{|{\cal I}|}{|{\cal U}|} \leq \frac{|{\cal I}|}{|{\cal I}_{\hat{\i}}|} \leq 1 + \frac{1}{h}\]
and the $(1+\epsilon)$-approximation is obtained setting $h = \lceil
1/\epsilon \rceil$.

For each $i \in \{0, \ldots, h\}$, the MIS can be solved exactly in
each block of the given partition in time $O(n
\ (hk)^{\frac{hk}{\omega}}\ \omega^{hk})$ and there is $O(n/hk)$
blocks. The overall running time is
\[O(h \ n^2\ \omega^{hk}\ (hk) ^{\frac{hk}{\omega}-1}).\]
Since $k=  \omega-1$ 
the running time is  
\[O(h \ n^2 \omega^{h(\omega-1)} (h (\omega-1))^{\frac{hk}{\omega}-1}).\]

%, which is exponential in $\omega$ and $h \approx 1/\epsilon$.
We have proved the following:

\begin{thm}
\label{new-ptas}
 There is a polynomial time approximation scheme for the {\rm MIS}
 problem in 2-dimensional LoS networks that for any $\epsilon>0$ returns a
 $(1+\epsilon)$-approximation in time $O(n^2\
 \omega^{\frac{\omega}{\epsilon}-1}\ (\frac{1}{\epsilon})^{\frac{1}{\epsilon}})$.
\end{thm}

\subsection{Arbitrary dimension $d > 2$} 

\begin{figure}[tb]
\includegraphics[scale=.35]{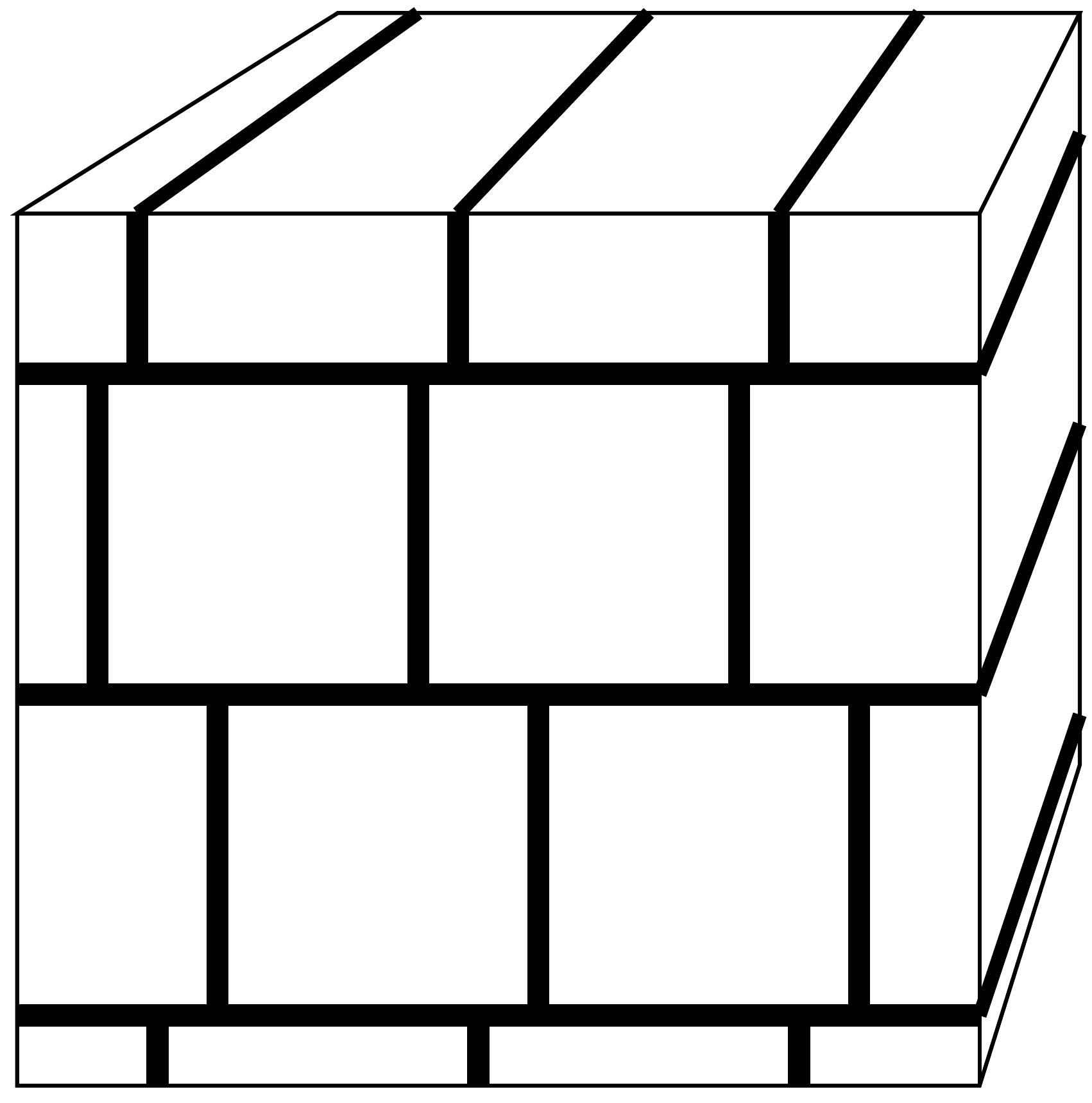}
\centering
\caption{\label{ptas-d} The whole LoS network partitioned into narrow
  LoS networks, for $d=3$. Note that the sizes of the narrow cubes may
vary. The black portions represent the vertices that are part of the
excluded strips.}
\end{figure}

A key feature of the PTAS
for $d=2$ is that the collection of strips $\cup_{i=0}^hV(\partial{\cal
  B}_{h,i})$ is a partition of the given graph vertex set. For $d>2$
the construction needs to be a bit careful.
Figure \ref{ptas-d} provides a diagrammatic picture of a possible PTAS
construction for $d=3$.

Let $b$ be an integer less than $d$. We say that a cube isomorphic to 
 \[\{1, \ldots, n\}  \times \ldots \times  \{1, \ldots, n\}  \times \underbrace{\{1, \ldots, x_1\} \times \ldots \times
  \{1, \ldots, x_b\}}_{b\ \mbox{\scriptsize terms}}\]
(where each of the $x_i \in \{1, \ldots, k\}$)
 is {\em a size $n$ cube narrow
  with respect to $b$ of its dimensions}.
In what follows a {\em size $n$, $b$-narrow}, $d$-dimensional LoS network is a LoS 
network whose nodes can be embedded in a size $n$ cube that is narrow with
respect to $b$ of its dimensions.
%, and $b(G)$ is
%the {\em width scale} of the graph $G$, the largest $b$ such that $G$
%is a $b$-narrow LoS networks.

\begin{claim}
\label{main-ptas}
Let $d$ and $b$ be fixed positive integers with $b<d$, and $n$ be an
arbitrary integer. For any
$\epsilon>0$, a $(1+\epsilon)^{d-b-1}$-approximation for the {\rm MIS} in a
size $n$, $b$-narrow, $d$-dimensional LoS network can be found in time
polynomial in $n$ but exponential in $\epsilon^{-1}$.
\end{claim}
\proof{
The claim can be proved by induction on
$d-b$. Let $G$ be an arbitrary size $n$, $b$-narrow, $d$-dimensional
LoS network.  
If $d-b \leq 1$ we can use the dynamic programming
strategy  described in Section \ref{algo} to solve the problem exactly.
For arbitrary $d-b$, without loss of generality assume that $G$ is
NOT narrow with respect to dimension one. Any part of $G$ that spans
the full length of the unrestricted dimensions, and is also
narrow with respect to dimension 1 is a size $n$, $b+1$ narrow, $d$
dimensional LoS network.
By the inductive hypothesis, the largest independent set of such
network 
can be approximated within $\left(1+\epsilon\right)^{d-b-2}$. For any
given $\epsilon$ let $h = \lceil 1/\epsilon \rceil$. Let $i \in \{0,
\ldots, h\}$. Given
$G$, we partition it into blocks containing all
nodes whose co-ordinates in $\mathbb{Z}^d_{n}$ have the first element
in the set $\{1, \ldots, ki\}$ or of the form
\[ki + (k+kh)(j-2) + l\]
where  $j \in \{2, \ldots, 2 + \lfloor(n-k i)/(k+kh)\rfloor\}$
and $l \in \{1, \ldots, k h\}$, except when $j$ takes its
largest value (in that case the largest value for $l$ is the
remainder of the integer division between $n-k i$ and $k+kh$). This defines a  partition ${\cal
  B}_{h,i}$ whose blocks are size $n$, $b+1$
narrow, $d$-dimensional LoS networks.
For one choice of $i$, which we denote again by $\hat{\i}$,  we must have 
\[|{\cal I} \cap V(\partial{\cal B}_{h,\hat{\i}})| \leq
\frac{|{\cal I}|}{h+1}\]
(where ${\cal I}$ is a maximum independent set of the whole network).
Therefore, reasoning like in the 2-dimensional case we have
\[|{\cal I}| \leq \left(1+\frac{1}{h}\right) |{\cal I}_{\hat{\i}}|\]
where ${\cal I}_{\hat{\i}}$ is the maximum  independent set of  ${\cal
  B}_{h,\hat{\i}}$. This independent set is  the union of
disjoint sets ${\cal
  I}_{\hat{\i}} \cap {\cal B}_{h,\hat{\i}}(j)$. 
By the inductive hypothesis there is an algorithm that finds a set $ {\cal
  U}_{\hat{\i}}(j)$ in the $j$th block of  ${\cal
  B}_{h,\hat{\i}}$ that satisfies
\[|{\cal
  I}_{\hat{\i}} \cap {\cal B}_{h,\hat{\i}}(j)| \leq
\left(1+\epsilon\right)^{d-b-2}  |{\cal
  U}_{\hat{\i}}(j)|\] 
Therefore
\[|{\cal I}| \leq \left(1+\frac{1}{h}\right)
\left(1+\epsilon\right)^{d-b-2}  \sum_j  |{\cal
  U}_{\hat{\i}}(j)|\]
and the claim follows.
} % proof

\begin{thm}
\label{new-ptas-d}
 There is a polynomial time approximation scheme for the {\rm MIS}
 problem in $d$-dimensional LoS networks.
\end{thm}
\proof{
For any fixed $\epsilon>0$, define $\epsilon' =
(1+\epsilon)^{\frac{1}{d-1}} - 1$. By Claim \ref{main-ptas} (with
$b=0$) there is
an algorithm that returns an independent set whose size is at least
$(1+\epsilon')^{1-d}$ that of a largest independent set in any given
 LoS network.
As to the running time, 
for each tuple $i_1, \ldots, i_{d-1}$, the MIS can be solved exactly in
each block of the given partition in time 
\[O\left(n\ (hk)^{\frac{(d-1)(hk)^{d-1}}{\omega}}\ \omega^{(hk)^{d-1}}\right)\]
and there is $O(n/(hk)^{d-1})$
blocks. The overall running time is
\[O\left(h^{d-1} \ \frac{n^2}{(hk)^{d-1}}\ \left(\omega\ (hk)
  ^{\frac{d-1}{\omega}}\right)^{(hk)^{d-1}}\right).\]
} % proof

\paragraph{Further Comments} It is perhaps instructive to compare
the algorithm presented in this section with the approximation scheme
described in \cite{DBLP:conf/caldam/SanghaZ17} for the MIS in general
$d$-dimensional LoS networks. The algorithm in that paper runs in time 
\[O(n^d (f_{d,\omega}(\epsilon))^{d
  (f_{d,\omega}(\epsilon))^d/\omega})\]
where 
\[f_{d,\omega}(\epsilon) = \frac{2 \cdot (d+1)!}{\omega}
\left(\frac{\omega}{\epsilon-\epsilon^2/2}\right)^{d+1}.\]
Since $f_{2,\omega}(\epsilon) = \frac{12 \cdot \omega^2}{(\epsilon-\epsilon^2/2)^3}$, 
for $d=2$,  the algorithm running time reduces to
\iffalse
\[O(n^2\left(\frac{12 \cdot
    \omega^2}{(\epsilon-\epsilon^2/2)^3}\right)^{\frac{2}{\omega}(\frac{12
    \cdot \omega^2}{(\epsilon-\epsilon^2/2)^3})^2}) = O(n^2\left(\frac{12 \cdot
    \omega^2}{(\epsilon-\epsilon^2/2)^3}\right)^{\frac{288
    \cdot \omega^3}{(\epsilon-\epsilon^2/2)^6}}) \]
and this is 
\fi
essentially
\[O(n^2\ \omega^{9 (4 \omega/\epsilon^2)^3}\
(\frac{12}{\epsilon-\epsilon^2/2})^{36 (2 \omega/\epsilon^2)^3})\]
which is much slower than the bound in Theorem \ref{new-ptas}, particularly
for small $\epsilon$ and moderate $\omega$.

The PTAS described in this section
is quite general and can be applied to several optimization problems
when the input is a 2-dimensional LoS network that is presented along
with its embedding in  $\mathbb{Z}^2_{n}$. In particular, simply
browsing through \cite{gary1979computers}
Vertex Cover, Min Dominating Set, Min Edge Dominating
Set,  Max Triangle Packing, Max $H$-matching, Max Tile Salvage can
all be solved to within $1+\epsilon$ of the optimum in a 2-dimensional
LoS network, if $\omega$ is a fixed constant independent of $n$.

%%%%%\newpage 
\section{Conclusions}
\label{sec:conclusions}

In this paper we study the maximum independent set problem on narrow
LoS networks. We propose an approach that  solves this
optimazion problem exactly in polynomial time on narrow LoS network,
presented with their $d$-dimensional embedding. We also describe how
such algorithm can be used as a
subroutine in a semi-online process that is guaranteed to return a
heuristic solution that is guaranteed to be only at most a factor
$1+\epsilon$ away from optimality, for any $\epsilon > 0$, in a
2-approximation algorithm for the MIS problem in arbitrary
$d$-dimensional networks, for fixed $\omega$ independent of $n$, and
in a PTAS for the 2-dimensional case.

We believe that the algorithmic ideas described here can be
generalized and applied to other  optimisation problems on LoS networks.

%\color{red}OPEN PROBS??\normalcolor

%%%%%\newpage

\end{document}